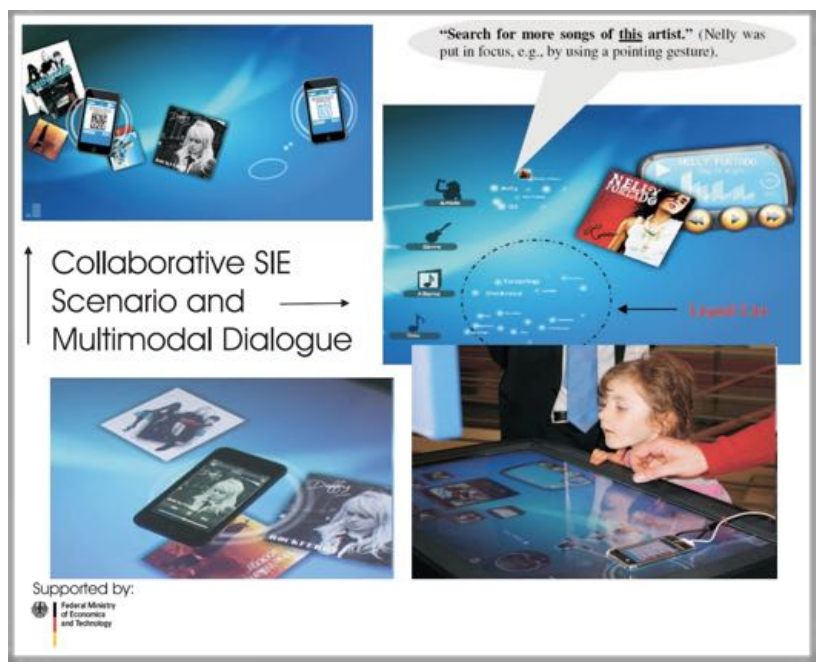

# Intelligent User Interfaces

will introduce you to the design and implementation of Intelligent User Interfaces (IUIs).

**ISMAR 2015 Tutorial on Intelligent User Interfaces**

Daniel Sonntag, German Research Centre for Artificial Intelligence (DFKI)

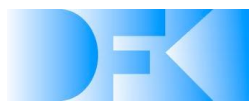

IUIs aim to incorporate intelligent automated capabilities in human computer interaction, where the net impact is a human-computer interaction that improves performance or usability in critical ways. It also involves designing and implementing an artificial intelligence (AI) component that effectively leverages human skills and capabilities, so that human performance with an application excels. IUIs embody capabilities that have traditionally been associated more strongly with humans than with computers: how to perceive, interpret, learn, use language, reason, plan, and decide.

## Motivation

Augmented and mixed reality are part of the input/output side of an intelligent user interface. There is a strong relationship between the intelligence in a system and the user interaction: first, intelligent processing is found in the user interface(s) of the system, and its purpose is to enable an effective, natural, or otherwise appropriate interaction of users with the system. For example, the system may support human-like communication methods such as speech or gesture; or it may adapt its style of interaction to individual users. Second, intelligent processing is found in the "backend" of the system, and its primary purpose is to serve some beneficial function such as performing actions partly autonomously on behalf of the users. In IUI, the relevance of the system's intelligence to interaction with users is in the fore.

In order to improve Augmented and Virtual Reality Applications, a deeper knowledge about the relationships between the aforementioned concepts/views (i.e., the intelligence in a system and the user interaction) would be highly beneficial. Mostly this demand refers to a deeper knowledge about the following major topics.

# Outline

- Design aspects of state-of-the-art intelligent user interfaces / interactive intelligent systems
- Semantic technologies: knowledge engineering in IUIs; user modelling
- Learning and automatic adaptation and planning in IUIs
- Input and output modalities (including the connection to VR/AR)
- Multimodal interaction, conversational agents, question answering
- Emotions and affective computing
- Applications and Projects, Human Computation, Collaborative Multimodality

This tutorial text is a compressed form of a new graduate lecture course at University of Kaiserslautern in 2013/2014/2015, see http://www.dfki.de/-sonntag/courses/WS14/IUI.html.

It is the worldwide first university lecture course in IUI. The presenter is a senior programme committee member of the IUI conference series.

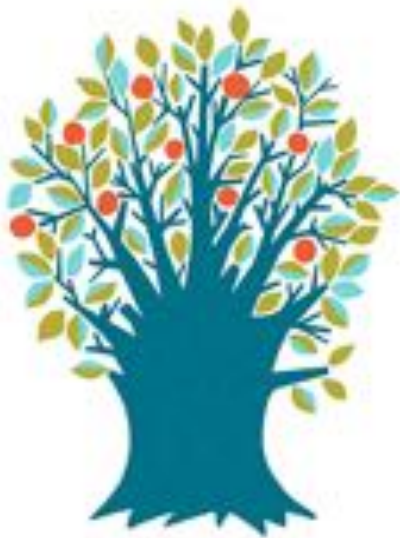

This tutorial text is a compressed form of a new graduate lecture course, see http://www.dfki.de/~sonntag/courses/WS14/IUI.html.

# IUI Community and Conferences

The IUI objectives are to increase productivity, decrease expenditures, and improve efficiency, effectiveness, and naturalness of interaction.

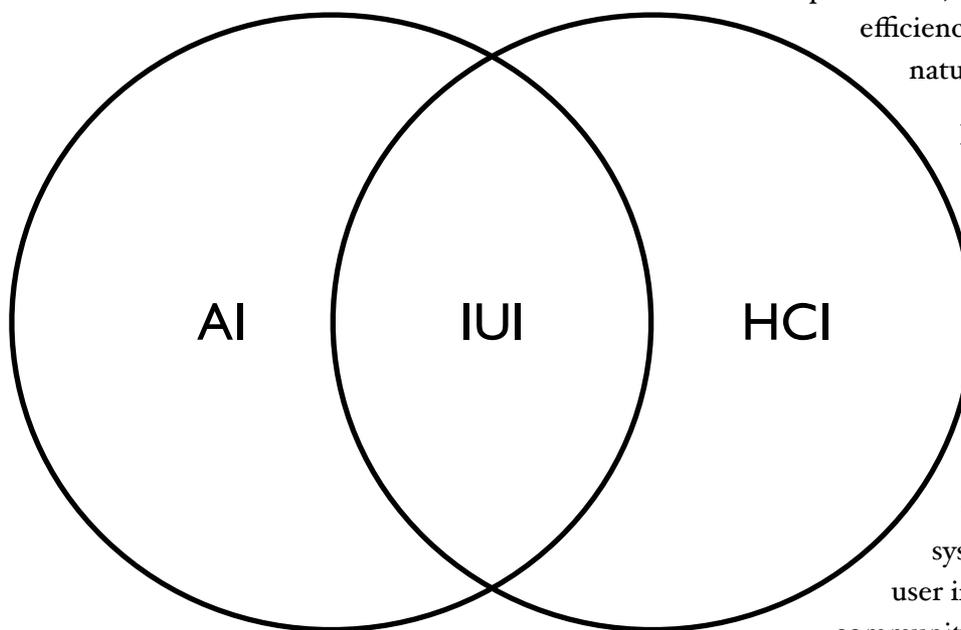

How? For example, by using knowledge representation, reasoning, ML, adaptation, and adaptivity. Examples include email filters, email response systems, spoken dialogue systems, and head-worn user interfaces. The IUI community is an interdisciplinary field and transcommunity of the sub-disciplines in artificial intelligence (AI) and human-computer interaction (HCI). The community works on the the design, realisation, and evaluation of interactive systems that exhibit some form of intelligence.

The main conference is the International Conference on Intelligent User Interfaces (ACM IUI), which is a SIGHCI conference. It should be mentioned that IUI is the only premier conference that focuses on the relationships between the intelligence in a system and the user interaction. Two cases have to be distinguished. (1) The intelligent processing is found in the user interface(s) of the system, and its purpose is to enable an effective, natural, or otherwise appropriate interaction of users with the system. For example, the system may support human-like communication methods such as speech or gesture; or it may adapt its style of interaction to individual users. Or, (2), the the intelligent processing is found in the "backend"of the

> **Relationships between the intelligence in a system and the user interaction**
>
> • The intelligent processing is used not directly in the system itself but in the process of designing, implementing, and/or testing the system. Hence, the system that the users interact with may not itself be an intelligent system.
>
> 21

system, and its primary purpose is to serve some beneficial function such as performing actions partly autonomously on behalf of the users. The relevance of the system's intelligence to interaction with users. A special case is when the intelligent processing is used not directly in the system.

There a many excellent conferences for the different cases. A quantitative overview of good conferences is given below, by Riedl and Jameson. Here, the term intelligent UI stands for techniques used to realise intelligent systems have their origins in AI—though in many cases a subfield has formed around a given type of technique (no longer primarily associated with AI). The canonical intelligent system includes a wide variety of capabilities, including sensing and perception, knowledge representation and reasoning, learning, creativity, planning, autonomous

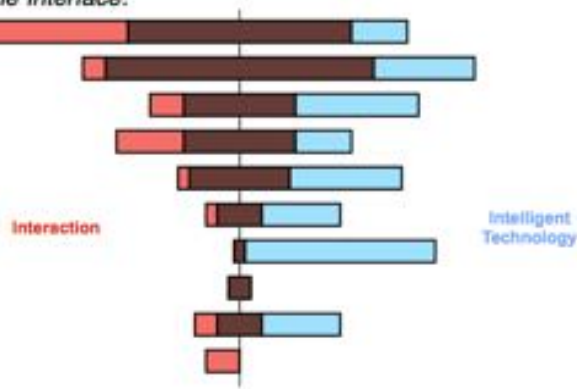
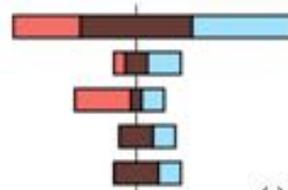

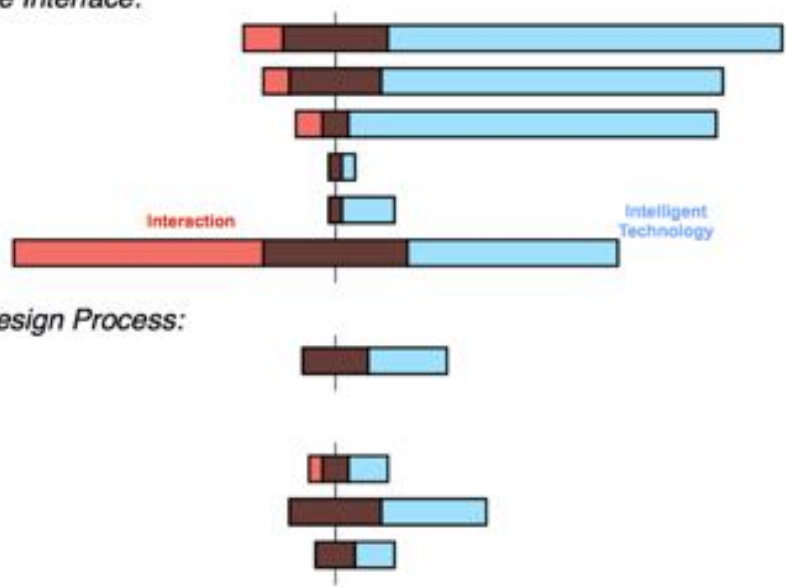

Some high-fidelity tools have been designed to support prototyping of new multimodal systems that support collaborative group interactions.

In order to improve Augmented and Virtual Reality Applications, a deeper knowledge about multimodal systems would be highly beneficial.

motion and manipulation, natural language processing, and social interaction.

Relationships between the intelligence in a system and the user interaction refer to our first case, the intelligent processing is found in the user interface(s) of the system, and its purpose is to enable an effective, natural, or otherwise appropriate interaction of users with the system. For example, the system may support human-like communication methods such as speech or gesture; or it may adapt its style of interaction to individual users. Systems where the intelligence lies mainly in the user interfaces have the following specific examples: systems with adaptive user interfaces are automatically adapted to the inferred capabilities or needs of the user; or multimodal systems that aim to enable more natural, human-like forms of input and output; systems with human-like virtual characters that enable the user to interact with a system in a way that is partly similar to human-human interaction; smart environments in which embedded objects interact intelligently with their users; personalised websites, in which the displayed content is adapted to the inferred interests of the user.

Systems where the intelligence lies mainly behind the user interfaces have the following implementations: recommender systems, which present products, documents, or other items that are expected to be of interest to the current user; systems that employ intelligent technology to support information retrieval; learning environments that offer learning assistance on the basis of assessments of each learner's capabilities and needs; interface agents that perform complex or repetitive tasks with some guidance from the user; situated assistance systems that monitor and support a user's daily activities; Systems for capturing knowledge from domain experts who are not knowledge engineers; games that make use of AI technology to create the opponents against which the human players play.

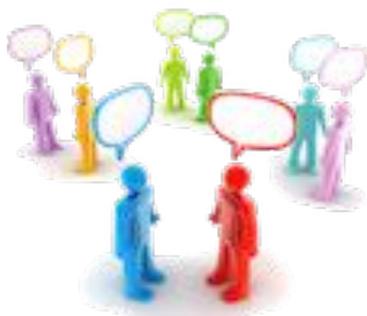

Systems where the intelligence lies mainly behind the user interfaces and systems where the intelligence lies mainly in the user interfaces are in the focus.

# Main questions and challenges

In what ways can artificial and human intelligence work together effectively?

Where does intelligent processing yield the greatest benefits for interaction, relative to other forms of computation?

What patterns of division of processing between the human and the intelligent system tend to be successful, and which less so?

# Challenges to usability and acceptance the incorporation of intelligence

The performance of an intelligent component may be fallible, leading to inappropriate interpretations or actions. Intelligent processing is often—though not inevitably—relatively difficult for users to predict, understand, and control. If they rely on relatively extensive information about the users, intelligent systems may raise certain types of privacy and security risks.

> **Relevant Research Areas**
> - In some other areas, the main contribution of the intelligence is to enhance communication between the system and users, in a way which **may or may not** be closely related to the system's main function. This is the contribution most commonly found in the areas of multimodal interaction, natural language processing, embodied conversational agents, computer graphics, and accessible computing.
>
> 13

An increased ability of a system to take over tasks that normally require thought and judgment can limit the breadth of experience and the responsibility of users.

How can we understand users' requirements for intelligent support in a particular context when the potential users have little idea of what sort of intelligent support is currently feasible?

How can we design an evaluation of a system that comprises measures of the performance of the intelligent algorithms, observations of users' behaviour, and the interpretation of users' subjective reactions?

There is a growing interest in and importance of IUI to answer these questions; and familiar examples of concrete suggestions and solutions include information retrieval systems that employ forms of intelligent processing, product recommender systems for e-commerce, telephone-based spoken dialogue systems; and household robots. No longer viewed as only embodying "artificial intelligence", there are also AI companions, i.e., (anthropomorphic) agents, and animated computer agents. See, for example:

http://www.youtube.com/watch?v=IhVu2hxm07E

http://www.youtube.com/watch?v=MaTfzYDZG8c

http://www.youtube.com/watch?v=_ySljCcnq4o

Other IUI solutions are often more application-specific and sometimes also more lightweight.

# Architecture of intelligent user interfaces

for implementing IUI's core areas: input analysis, output generation, user- and discourse-adapted interaction, agent-based interaction, model-based interface design, and intelligent interface evaluation. A given system may involve intelligence on more than one level. For example, an AI-based game may offer multimodal interaction with the user as well as intelligent game playing on the part of the game's characters. More complex architectures are needed to account for such a complex behaviour; often they take users models, discourse models, domain models, and task models, and media models into account. In addition, the input processing features a comprehensive media analysis competency and **automated** media analysis, design, and interaction management. Also, specific user modelling approaches (Kobsa and Wahlster 89, Horvitz 97) and planning approaches (Allen and Hendler 90) can have a major role to play. The analysis of input is based on multimodal input (e.g., spoken, typed, and hand written language; gestures, including hand, eye, and body states and motion). The generation (planning or realisation according to, e.g., user models) is the step that co-ordinates the output which is presented to the user.

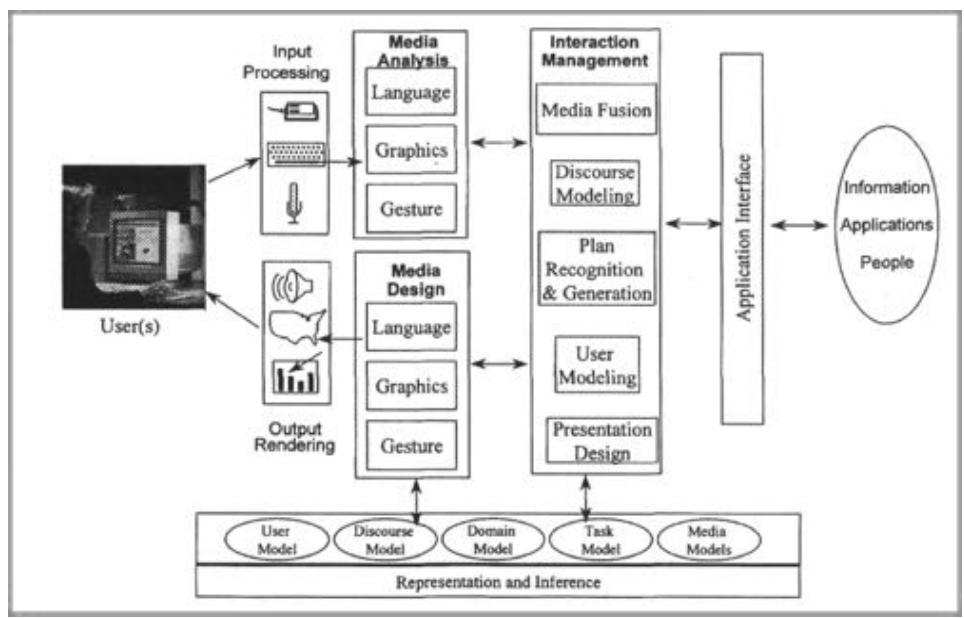

It can be said that the modelling of the user, discourse, task, and situation and interaction management, including possible tailoring of interaction to the user, task, and/or situation, is the main groundwork and enabler for **adaptive and adaptable user interface**s.

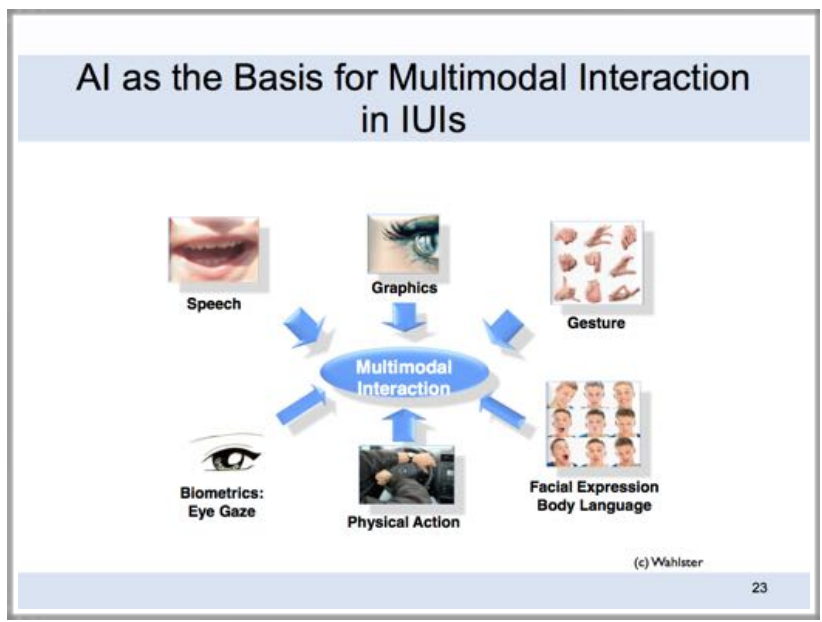

An important additional source of information for new mixed reality interfaces is the new handbook on multimodal user interfaces by Oviatt and Cohen, "The Paradigm Shift to Multimodality." This book is based on the theory of multimodal interaction, where AI is the basis for multimodal interaction in IUIs and explains the foundation of human-centred multimodal interaction and interface design, based on the cognitive and neurosciences, as well as the major benefits of multimodal interfaces for human cognition and performance. It describes the data-intensive methodologies used to envision, prototype, and evaluate new multi-modal interfaces. From a system development viewpoint, this book outlines major approaches for multimodal signal processing, fusion, architectures, and techniques for robustly interpreting users' meaning. The ISMAR community benefits most from a detailed review of techniques for robustly interpreting users' meaning from multiple input sources. This knowledge will help in making mixed reality reality interface more responsive in the future by automatically interpreting multimodal input in real-time and providing mixed reality visualisations in real-time.

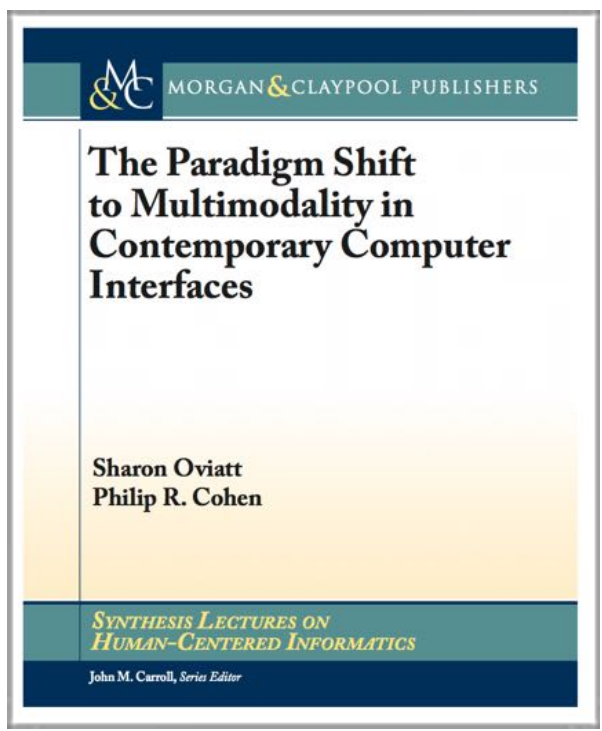

In addition, the Intelligent User Interfaces department at DFKI develops the basics for multi-modal human-computer interaction and personalised dialogue systems integrating speech, gestures, and facial expressions with physical interaction. In this process, user, task, and domain models are used to design dialogue patterns that are as natural and robust as possible—even in group discussions or loud environments.

For example, by integrating virtual characters, even emotions and socially interactive behaviour can be achieved as an output (in addition to the multimodal input fusion aspect that includes emotion processing in the sense of human computing). A major focus lies mobile user interfaces to location-based and context-sensitive services especially for specific application domains such as medicine. In addition to the intuitive access to the Internet of Services and Things in the context of the semantic web, the DFKI department also studies barrier-free access to instrumented environments and networked worlds for seniors. Significant for this kind of IUI research and the DFKI department is the interdisciplinary method of operation especially the cooperation of computer scientists with computational linguists and cognitive psychologists, as well as the empirical and ergonomic evaluation of spoken dialogue systems and multimodal user interfaces.

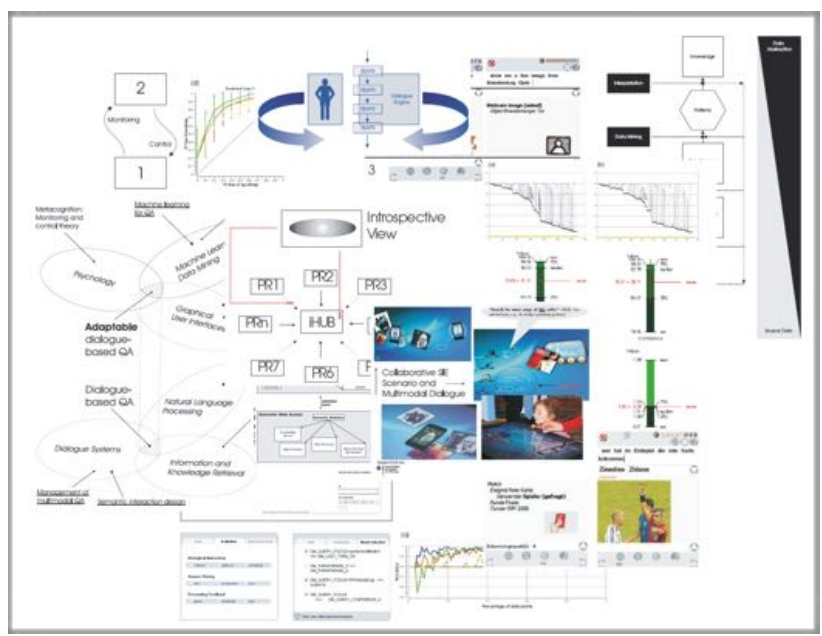

The research programme THESEUS of the Federal Ministry of Economics and Technology aimed at developing a new internet based knowledge infrastructure in order to better deploy and make use of the information contained on the internet. In the context of innovative basic technologies, the DFKI intelligent user interface research department has been responsible for the work package "situation-aware dialogue shell" in which multimodal interfaces for dialogue interaction between man and machine are being implemented. Thus, users can formulate their questions intuitively and refine them during the dialogue with the system.

DFKI's main publication within the research pre-project SmatyWeb is the book "Ontologies and Adaptivity in Dialogue for Question Answering", see http://www.dfki.de/~sonntag/Ontologies_and_Adaptivity_in_Dialogue_for_Question_Answering.html. The appeal of being able to ask a question to a mobile internet terminal and receive an answer immediately has been renewed by the broad availability of information on the Web. Ideally, a spoken dialogue system that uses the Web as its knowledge base would be able to answer a broad range of questions. A new generation of natural language dialogue systems is emerging that transforms traditional keyword search engines into semantic answering machines by providing exact and concise answers formulated in natural language instead of today's long lists of document references, which the user has to check by himself for relevant answers.

This book presents the anatomy of the fully operational SmartWeb system (funded by the German Federal Ministry of Education and Research with grants totaling 14 million euros) that provides not only an open-domain question answering machine but a multimodal web service interface for coherent dialogue, where questions and commands are interpreted according to the context of the previous conversation. One of the key IUI innovations described in this book is the ability of the system to learn how to predict the probability that it can answer a complex user query in a given time interval.

Within the use case THESEUS TEXO researchers then developed an infrastructure for new web-based cross-company services that rely on Service Oriented Architecture (SOA). In this context the IUI research department creates a mobile, multimodal interface for the access to Business Webs.

Eye Tracking has many applications in active or passive mode.

In addition, the Innovation Alliance Digital Product Memory, being part of the promotion programme IKT2020, has researched on the semantic interoperability of different product memories for ubiquitous, multimodal access to Digital Product Memories with partners from the fields of research, trade, logistics, the pharmaceutical industry and automobile industry - lead by our research department. Work on the development of open standards is actively continued in the W3C Object Memory Modelling Incubator Group . SemProM has developed the basics for the next generation of mobile, embedded and function-based elements for the semantic internet communication between everyday objects. Thereby, capacities of such "intelligent" products go far beyond the mere identification function of current RFID labelling. They are able to evaluate data of different embedded sensors (e.g., temperature, lights, humidity, speed, acceleration, position) and to register all relevant product and process data at the same time. In terms of the "Internet of Things" this information can be actively exchanged with their environment, users and others products. Internet of Things is mainly about connected devices embedded in our everyday environment. Typically, 'interaction' in the context of IoT means interfaces which allow people to either monitor or configure IoT devices. Some examples include mobile applications and embedded touch screens for control of various functions (e.g., heating, lights, and energy efficiency) in environments such as homes and offices. In some cases, humans are an explicit part of the scenario, such as in those cases where people are monitored (e.g., children and elderly) by IoT devices. Interaction in such applications is still quite straightforward, mainly consisting of traditional graphical interfaces, which often leads to clumsy co-existence of human and IoT devices. Thus, there is a need to investigate what kinds of interaction techniques could provide IoT to be more human oriented, what is the role of automation and interaction, and how human originated data can be used in IoT.

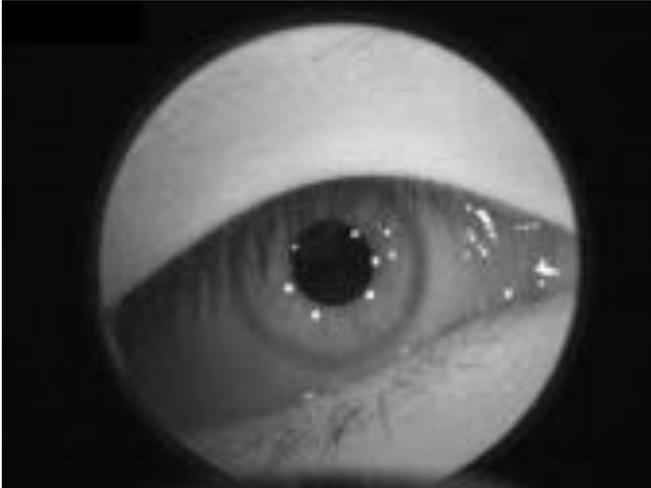

The joint project RES-COM (resource conservation by context-activated machine-to-machine communication) will play an important role towards resource-conserving production and service offerings in the context of recommendations presented by the German Industry-Science Research Alliance concerning the project Industry 4.0 for the future.

We follow up on these so-called carrier projects in the EIT Digital framework, where one of the light-house project is called "Cyber-physical Systems for Smart Factories". This H2020 project of the European communion (project lead Daniel Sonntag, DFKI) combines Industry 4.0 technology with intelligent user interface technology for human-robot collaboration. See http://dfki.de/smartfactories/.

The over-arching challenge to address is to combine CPS safety and performance. While addressing safety challenges, the outcomes include models of the behaviour of loops with human operators, in particular how to ensure safety. Because humans are unpredictable, a special outcome is a model which account for humans as producing anomalies by reacting to predictable maintenance tasks and unpredictable events.

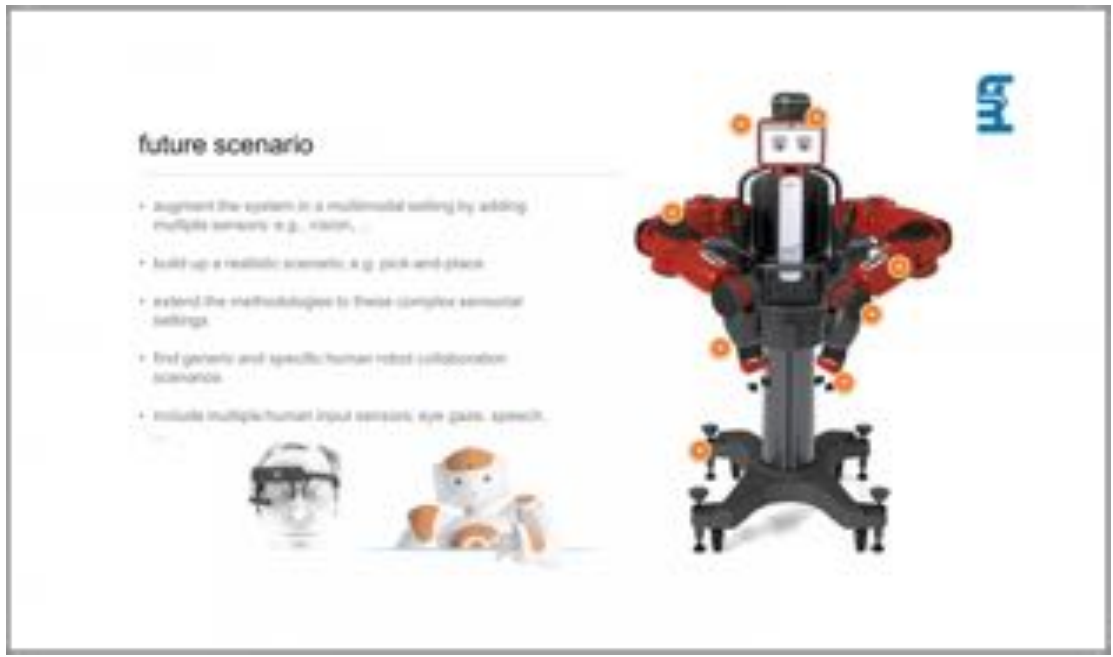

Technical advancements:

- GPU-based deep learning machine learning infrastructure for anomaly treatment and data mining

- RFID based product memory infrastructure

- Smart factories knowledge portal infrastructure for an anomaly instance base (knowledge management)

- Model based prediction with anomaly detection followed by machine learning and real time verification

- Intelligent user interfaces for expert knowledge acquisition, human behaviour input and human-robot interaction by using, e.g., vision sensors.

These components are built into an architecture and are to be extended with the characterisation of a human collaborator who is also "in the loop" and may also exhibit anomalous behaviour. Cyber-physical systems are implemented in human

http://www.dfki.de/wwdata/Publications/Introduction_to_intelligent_User_Interfaces.pdf

INTRODUCTION

# Intelligent User Interfaces: An Introduction

This introduction describes the need for intelligent user interfaces (IUIs), specifies the intended purpose and use of this collection, outlines the collection's scope, and defines basic terminology used in the field. After outlining the theoretical foundations of intelligent user interfaces, this introductory section describes the current state of the art and summarizes the structure and contents of this collection, which addresses some remaining fundamental problems in the field.

### 1. MOTIVATION

The explosion of available materials on corporate, national, and global information networks is driving the need for more effective, efficient, and natural interfaces to support access to information, applications, and people. This is exacerbated by the increasing complexity of systems, the shrinking of task time lines, and the need to reduce the cost of application and interface development. Fortunately, the basic infrastructure for advanced multimedia user interfaces is rapidly appearing or already available. In addition to traditional public telephone networks, cable, fiber-optic, wireless and satellite communications are rapidly evolving with the aim of serving many simultaneous users through a great variety of multimedia communications (e.g., video, audio, text, data). Rapidly advancing microprocessor and storage capabilities, coupled with multimedia input and output devices integrated into workstations and portable machines, provide a dizzying array of potential for personal and personalized multimedia interaction.

Interface technology has advanced from initial command line interfaces to the established use of direct manipulation or WIMP (windows, icons, menus, and pointing) interfaces in nearly all applications. Even some of the first computing systems incorporated graphical displays and light pens as pointing devices (Everett et al. 1957). The next generation of interfaces, often called "intelligent," will provide a number of additional benefits to users, including adaptivity, context sensitivity, and task assistance. As with traditional interfaces, principled intelligent interfaces should be learnable, usable, and transparent. In contrast, however, intelligent user interfaces promise to provide additional benefits to users that can enhance interaction, such as:

- Comprehension of possibly imprecise, ambiguous, and/or partial multimodal input
- Generation of coordinated, cohesive, and coherent multimodal presentations
- Semi- or fully automated completion of delegated tasks
- Management of the interaction (e.g., task completion, tailoring interaction styles, adapting the interface) by representing, reasoning, and exploiting models of the user, domain, task, and context

Lieberman H (2009) User interface goals, AI opportunities.
AI Mag 30(4):16–22

Articles

# User Interface Goals, AI Opportunities

Henry Lieberman

■ This is an opinion piece about the relationship between the fields of human-computer interaction (HCI) and artificial intelligence (AI). The ultimate goal of both fields is to make user interfaces more effective and easier for people to use. But historically, researchers have disagreed about whether "intelligence" or "direct manipulation" is the better route to achieving this. There is an unjustified perception in HCI that AI is unreliable. There is an unjustified perception in AI that interfaces are merely cosmetic. This disagreement is counterproductive.

This article argues that AI's goals of intelligent interfaces would benefit enormously by the user-centered design and testing principles of HCI. It argues that HCI's stated goals of meeting the needs of users and interacting in natural ways would be best served by application of AI. Peace.

*I*s AI antithetical to good user interface design? From the earliest times in the development of computers, activities in human-computer interaction (HCI) and AI have been intertwined. But as subfields of computer science, HCI and AI have always had a love-hate relationship. The goal of HCI is to make computers easier to use and more helpful to their users. The goal of artificial intelligence is to model human thinking and to embody those mechanisms in computers. How are these goals related?

Some in HCI have seen these goals sometimes in opposition. They worry that the heuristic nature of many AI algorithms will lead to unreliability in the interface. They worry that AI's emphasis on mimicking human decision-making functions might usurp the decision-making prerogative of the human user.

These concerns are not completely without merit. There are certainly many examples of failed attempts to prematurely foist AI on the public. These attempts gave AI a bad name, at least at the time.

But so too have there been failed attempts to popularize new HCI approaches. The first commercial versions of window systems, such as the Xerox Star and early versions of Microsoft Windows, weren't well accepted at the time of their introduction. Later design iterations of window systems, such as the Macintosh and Windows 3.0, finally achieved success. Key was that these early failures did not lead their developers to conclude window systems were a bad idea. Researchers shouldn't construe these (perceived) AI failures as a refutation of the idea of AI in interfaces.

Modern PDA, smartphone, and tablet computers are now beginning to have quite usable handwriting recognition. Voice recognition is being increasingly employed on phones, and even in the noisy environment of cars. Animated agents, more polite, less intrusive, and better thought out, might also make a

16   AI MAGAZINE   

*Moving from traditional interfaces toward interfaces offering users greater expressive power, naturalness, and portability.*

# Ten Myths of Multimodal Interaction

**Sharon Oviatt**

Multimodal systems process combined natural input modes—such as speech, pen, touch, hand gestures, eye gaze, and head and body movements—in a coordinated manner with multimedia system output. These systems represent a new direction for computing that draws from novel input and output technologies currently becoming available. Since the appearance of Bolt's [1] "Put That There" demonstration system, which processed speech in parallel with manual pointing, a variety of multimodal systems has emerged. Some rudimentary ones process speech combined with mouse pointing, such as the early CUBRICON system [8]. Others recognize speech while determining the location of pointing from users' manual gestures or gaze [5].

Recent multimodal systems now recognize a broader range of signal integrations, which are no longer limited to the simple point-and-speak combinations handled by earlier systems. For example, the Quickset system integrates speech with pen input that includes drawn graphics, symbols, gestures, and pointing. It uses a semantic unification process to combine the meaningful multimodal information carried by two input signals, both of which are rich and multidimensional. Quickset also uses a multi-agent architecture and runs on a handheld PC [3]. Figure 1 illustrates Quickset's response to the multimodal command "Airstrips... facing this way, facing this way, and facing this way," which was spoken while the user drew arrows placing three airstrips in correct orientation on a map.

Multimodal systems represent a research-level paradigm shift away from conventional windows-icons-menus-pointers (WIMP) interfaces toward providing users with greater expressive power, naturalness, flexibility, and portability. Well-designed multimodal systems integrate complementary modalities to yield a highly synergistic blend in which the strengths of each mode are capitalized upon and used to overcome weaknesses in the other. Such systems potentially can function more robustly than unimodal systems that involve a single recognition-based technology such as speech, pen, or vision.

FACHBEITRAG

# Collaborative Multimodality

**Daniel Sonntag**

www.dfki.de/-sonntag/collaborative-multimodality.pdf

**Abstract** This essay is a personal reflection from an Artificial Intelligence (AI) perspective on the term HCI. Especially for the transfer of AI-based HCI into industrial environments, we survey existing approaches and examine how AI helps to solve fundamental problems of HCI technology. The user and the system must have a collaborative goal. The concept of *collaborative multimodality* could serve as the missing link between traditional HCI and intuitive human-centred designs in the form of, e.g., natural language interfaces or intelligent environments. Examples are provided in the medical imaging domain.

**Keywords** AI methods · Multimodal interaction · Dialogue systems · Collaboration

## 1 Introduction

The term Human-Computer Interaction (HCI) confuses some researchers and practitioners. Many think of HCI as including diverse areas of traditional graphical and web user interfaces, others rather think of new multimodal input and output devices, tangible user interfaces, virtual and augmented reality, intelligent environments, and/or interfaces in the ubiquitous computing paradigm. Whereas the supporters of the traditional HCI view have a strong motivation and justification in the desktop-based ubiquitousness of traditional computer terminals with computer screens, psychological analysis background, and integral evaluation methods, the new AI-based technologies can impress with intuitive human-centred designs. (It should be noted that human-centred designs do not necessarily improve the usability of an HCI, especially in industrial environments.)

HCI is the business of designing user interfaces that people can work well with. Hence, it is an area of research, design, and application, which combines all the aforementioned diverse areas. There is a great variety in these partly overlapping areas which are all involved in this business.

In this article, the first goal is to give an overview of the AI-based HCI techniques which include multimodal interaction. We think that in the future, AI will have a great influence on multimodal interaction systems. Therefore, we will begin by articulating the key issues of the concept of multimodal interfaces in the sense of a combination of traditional, screen-based HCI techniques and interfaces with new (mobile) multimodal input and output devices and interaction forms. Goebel and Williams have commented on the goal to stitch together the breadth of disciplines impinging on AI [8]; following their idea, we try to stitch together the breadth of disciplines impinging on AI-based multimodal HCI. For this purpose, we will introduce the notion of *collaborative multimodality*, which could serve as the missing link between traditional HCI and intuitive human-centred designs.

The second goal is to give a summary of the various different approaches taken by ourselves and other participants in the research field of multimodal dialogue-based HCI for prototyping industry-relevant applications of intelligent user interfaces (IUIs). We think that collaborative multimodality as introduced here represents one of the major usability requirements. AI methods such as sensory input interpretation and explicit models of the discourse of the interaction and the domain of interest are employed to create an argument in favour of the hypothesis that the emergence of a complex, collaborative multimodal behaviour is what best describes an intelligent user interface. To prove this,

D. Sonntag (✉)
German Research Center for AI (DFKI), Stuhlsatzenhausweg 3, 66123 Saarbruecken, Germany
e-mail: sonntag@dfki.de

# Kognit: Intelligent Cognitive Enhancement Technology by Cognitive Models and Mixed Reality for Dementia Patients

**Daniel Sonntag**
German Research Center for Artificial Intelligence (DFKI)
Stuhlsatzenhausweg 3
Saarbruecken, Germany

**Abstract**

With advancements in technology, smartphones can already serve as memory aids. Electronic calendars are of great use in time-based memory tasks. In this project, we enter the mixed reality realm for helping dementia patients. Dementia is a general term for a decline in mental ability severe enough to interfere with daily life. Memory loss is an example. Here, mixed reality refers to the merging of real and virtual worlds to produce new episodic memory visualisations where physical and digital objects co-exist and interact in real-time. Cognitive models are approximations of a patient's mental abilities and limitations involving conscious mental activities (such as thinking, understanding, learning, and remembering). External representations of episodic memory help patients and caregivers coordinate their actions with one another. We advocate distributed cognition, which involves the coordination between individuals, artefacts and the environment, in four main implementations of artificial intelligence technology in the Kognit storyboard: (1) speech dialogue and episodic memory retrieval; (2) monitoring medication management and tracking an elder's behaviour (e.g., drinking water); (3) eye tracking and modelling cognitive abilities; and (4) serious game development towards active memory training. We discuss the storyboard, use cases and usage scenarios, and some implementation details of cognitive models and mixed reality hardware for the patient. The purpose of future studies is to determine the extent to which cognitive enhancement technology can be used to decrease caregiver burden.

## Introduction

Dementia is a mental disorder that is associated with a progressive decline in mental functions and abilities. Memory, thinking, language, understanding, and judgement are affected. While many older adults will remain healthy and productive, overall this segment of the population is subject to cognitive impairment at higher rates than younger people. There are two important directions of research: the use of AI to elders with dementia and the design of advanced assistive technology. We combine those two to form a new research field: *intelligent cognitive enhancement technology*. Cognitive assistance can be characterised as therapeutic, aimed at correcting a specific pathology or defect. Some of our implementations fall into this category. Enhancement however is an intervention that improves a subsystem in some way other than repairing something that is broken or dysfunctioning. This is of particular interest because new intelligent user interfaces in the form of vision-based wearable devices give patients, and humans in general, new effective cognitive abilities (most notably in the area of information retrieval.) "In practice, the distinction between therapy and enhancement is often difficult to discern, and it could be argued that it lacks practical significance." (Bostrom and Sandberg 2009)

Intelligent cognitive assistance and enhancement technologies may enable older adults to live independently for longer periods of time. In 2014, the Alzheimer's Association documented that approximately 10-20% of the population over 65 years of age suffer from mild cognitive impairment (MCI); in 2013, Americans provided billion hours of unpaid care to people with Alzheimer's disease and other dementias.[1]

The envisioned benefit is to improve independent and self-determined living. As institutionalisation has an enormous financial cost, intelligent cognitive enhancement technologies may also help to reduce healthcare system costs, while at the same time provide relief and more time for caregivers and family members. In Kognit, we focus on new computer-based technologies on the horizon that offer help for patients directly by supporting an independent and self-determined living and, indirectly, in the caregiving of dementia patients, to provide relief and more time for family members. Cognitive models and mixed reality should result in new cognitive enhancement technology (figure 1).

Our research about cognitive enhancement technology is supported by a number of clinical research and related works about memories of daily life activities, a design case study for Alzheimer's disease (Cohene et al. 2007), reality orientation for geriatric patients (Taulbee, Lucille R and Folsom, James C 1966), using validation techniques with people living with dementia, computer based technology and caring for older adults (Spry Foundation 2003), and non-pharmaceutical treatment options for dementia (Douglas, James, and Ballard 2004). The focus is on cognitive assistants (Cogs) interfaces that include sensor interpretation, activity recognition, pro-active episodic memory, and environments! This represents a new "cyber-physical" form of IUIs. The software/hardware outcome package consists of anomaly controllers for smart factories. We focus on both open and closed-loop controllers in the robot domain and reporting/maintenance domain in manufacturing. Similar medical CPS examples with multimodal interaction have been implemented in ERmed, see http://www.dfki.de/RadSpeech/ERmed.

---
[1]See http://www.alz.org/downloads/facts_figures_2014.pdf

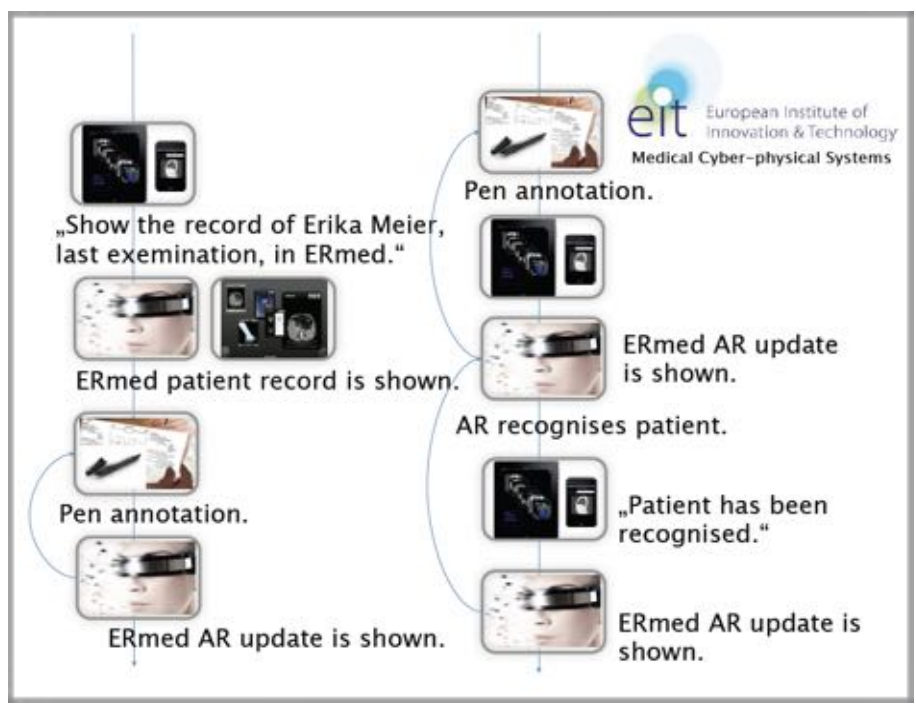

The four introductory Papers for IUI—and one recent application paper—you have just seen should provide you with a additional information on the theoretical foundations of intelligent user interfaces, this introductory section describes the current state of the art. A number of resources for teachers, students, and researchers contain additional information about this subject area. Some of these selected resources can be found in the last section of this tutorial text.

# How to read an IUI research paper

Reading IUI research papers effectively is challenging. These papers are often written in a very condensed style because of page limitations and the intended audience, which is assumed to already know the area well. Moreover, the reasons for writing the paper may be different than the reasons the paper has been assigned, meaning you have to work harder to find the content that you are interested in.

The first thing to understand is that the research papers you will read have been published in different kinds of places. Some papers have been published in the proceedings of a conference, for example at **IUI**, a SIGART / SIGCHI conference submission.

These papers have been peer-reviewed but are typically strictly limited in length to about 10-12 pages. Other papers have been published in archival journals, such as **TiiS**. These papers have also been peer-reviewed but there are typically not any length restrictions, so they generally include more detail in every dimension. Some papers are technical reports. These are not peer-reviewed. In areas related to Computer Science, you may find that there is first a technical report version which is then later published in a conference or a journal. If a paper appears both in conference and journal form, the journal version is typically a later, expanded and revised version of the work.

To develop an effective reading style for research papers, it can help to know what you should get out of the paper, and where that information is located in the paper. Typically, the introduction will state not only the motivations behind the work, but also outline the solution. Often this may be all the expert requires from the paper. The body of the paper states the authors' solution to the problem in detail, and should also describe a detailed evaluation of the solution in terms of arguments or an empirical evaluation (case study, experiment, etc.). Finally, the paper will conclude with a recap, including a discussion of the primary contributions. A paper will also discuss related work to some degree.

The questions you want to have answered by reading a paper in order to communicate effectively in a professional environment & work in IUI group projects, review and evaluate IUIs (to a certain extend), recognise the need to keep up to date with developments in IUI, participate professionally in industrial research and development (after taking related information science courses) are the following:

**What are motivations for this work?** For a research paper, there is an expectation that a solution to a problem has been found that no one else has published in the literature. The paper should describe why the problem is important and why it does not have a trivial solution; that is, why a new solution may be required. Implicitly there is implication that previous solutions to the problem are inadequate. Occasionally an author will fail to state either point, making your job much more difficult.

**What is the proposed solution?** This is also called the hypothesis or idea. There should also be an argument about why the solution solves the problem better than previous solutions. There should also be a discussion about how the solution is achieved (designed and implemented) or is at least achievable.

**How is the proposed solution evaluated?** An idea alone is usually not adequate for publication of a research paper. What argument and/or experiment is made to make a case for the value of the ideas? What benefits or problems are identified? Are they convincing?

**How does the proposed solution relate to other proposed solutions?** If this is an important problem, other researchers have probably also tried to solve it. How do the alternative approaches compare to this one? Has this work already been done? Don't limit yourself to the related work references cited by the authors.

**What are the contributions?** The contributions in a paper may be many and varied. Ideas, software, experimental techniques, and area survey are a few key possibilities.

**What are future directions for this research?** Not only what future directions do the authors identify, but what ideas did you come up with while reading the paper?

# Research Paper Reading List

## More conceptual

1	Wahlster W (1998) Intelligent User Interfaces: An Introduction. In   Maybury, Wahlster: Readings in Intelligent User Interfaces

2	Lieberman H (2009) User interface goals, AI opportunities. AI Mag 30(4):16–22

3	Horvitz E (1999) Uncertainty, action, and interaction: In pursuit of mixed-initiative computing. IEEE Intell Syst 14:17–20

4	Maybury MT, Stock O, Wahlster W (2006) Intelligent interactive entertainment grand challenges. In: Proc of IEEE intelligent systems pp. 14–18

5	Rich C, Sidner CL, Lesh N (2001) Collagen: applying collaborative discourse theory to human-computer interaction. AI Mag 22(4):15–26

6	Oviatt S (1999) Ten myths of multimodal interaction. Commun ACM 42(11):74–81

7	Sonntag, D (2012) Collaborative Multimodality, KI - German Journal on Artificial Intelligence 26 (2):161–168

8	McGuiness, D. (2004) Question Answering on the Semantic Web,    IEEE Intelligent Systems 19(1):82–85

9	Lieberman H,   Liu H,   Singh P,   Barry B (2004), Beating common sense into interactive applications, AI Magazine 25(4):63–76. AAAI Press.

10	Horvitz E,   Kadie C,   Paek T, and   Hovel D (2003) Models of attention in computing and communication: from principles to applications. Commun. ACM 46, (3):52–59.

## More technical

1	Sarwar B,   Karypis G,   Konstan J,   Riedl J (2001) Item-based collaborative filtering recommendation algorithms. WWW 2001: 285-295

# Books and Proceedings